%% file: main.tex
\documentclass{llncs}
\usepackage{mathtools}
\usepackage{amssymb,amsmath}
\usepackage{bm}
\usepackage{array}
\usepackage{algpseudocode}
\usepackage{tikz}
\usepackage{url}
\usepackage{enumerate}
\usepackage{microtype}
\usepackage{paralist}
\usepackage{comment}
\usepackage{wrapfig}
\usepackage{stmaryrd}
\usepackage{todonotes}
\usepackage{thm-restate}

\usepackage{bbm}
\usepackage{macros}
\usepackage{graphicx}

\usepackage[linesnumbered,ruled,vlined,noend]{algorithm2e}
\SetKwData{KwInput}{Input:}{}
\SetKwInput{kwInit}{Init}

\usepackage{hyperref}

\AtEndEnvironment{proof}{\phantom{}\qed}
\usepackage{cleveref}
\crefname{algorithm}{Algorithm}{Algorithm}
\Crefname{algorithm}{Algorithm}{Algorithms}
\crefname{proposition}{Prop.}{Prop.}
\Crefname{proposition}{Proposition}{Propositions}
\crefname{section}{Sect.}{Sect.}
\Crefname{section}{Section}{Sections}
\crefname{figure}{Fig.}{Fig.}
\Crefname{figure}{Figure}{Figures}
\crefname{example}{Ex.}{Ex.}
\Crefname{example}{Example}{Examples}

\usepackage{subcaption}
\usetikzlibrary{automata,calc}
\usetikzlibrary{arrows,shapes,decorations,decorations.pathreplacing,calligraphy}

\title{Distributed Monitoring of Timed Properties}

\author{L\'eo Henry\inst{1} \and
  Thierry J\'eron\inst{2} \and
  Nicolas Markey\inst{2} \and
  Victor Roussanaly\inst{3}}

\institute{University College London, UK \and
  IRISA, CNRS \& Inria \& Univ. Rennes, France \and
  Universit\'e de Lorraine, France}

\begin{document}

\maketitle

\begin{abstract}
\looseness=-1
  In formal verification, runtime
  monitoring consists of observing the execution of a system in order
  to decide as quickly as possible whether or not it satisfies a given
  property.  
  We~consider monitoring in a distributed
setting, for properties given as reachability timed automata.  In~such
a~setting, the~system is made of several components, each equipped with its
own local clock and monitor. The~monitors observe events occurring on
their associated component, and receive timestamped events from other
monitors through FIFO channels.  Since clocks are local, they cannot
be perfectly synchronized, 
resulting in imprecise timestamps. 
Consequently, they~must be seen as intervals, leading
monitors to consider possible reorderings of events.
In~this context, each~monitor aims to provide, 
as early as possible,
a~verdict on the property it~is~monitoring, based~on its potentially
incomplete and imprecise knowledge of the current execution.

In this paper, we propose an on-line monitoring algorithm for
timed properties, robust to time imprecision and partial information
from distant components.  We~first identify the date at which a
monitor can safely compute a verdict based on received events.
We~then propose a monitoring algorithm that updates this date
when new information arrives,
maintains the current set of states in which the
property can reside, and updates its verdict accordingly.
\end{abstract}

\input{intro}

\input{preliminaries}

\input{base}
\input{conclu}

\paragraph*{Acknowledgements.}
This work was partially supported by the EPSRC Standard Grant CLeVer
(EP/S028641/1).


\input{main.bbl}
\end{document}

%% file: intro.tex
\section{Introduction}
\paragraph{Runtime verification.}
Formal verification is a branch of computer science that aims to
check whether computer systems satisfy their requirements. This includes
in particular model checking~\cite{hbmc,BK08} and deductive
verification~\cite{Hoa69,Fil11}, which work offline and check properties of all
behaviours of the system under study.  In~contrast, runtime
verification~\cite{LS09,BF18} is a set of efficient techniques to
monitor the behaviour of a \emph{running} system, and to
detect, as early as possible during the execution, whether some properties
are satisfied or violated.
This domain of formal verification has been extensively studied over
the last 25 years, and is now mature enough to be applied in various
application domains.

As real-life systems typically comprise several connected components,
runtime verification techniques must be adapted to handle
situations where
the behaviours of the components are only partially known and may be imprecise:
each component is equipped with a monitor, and monitors exchange
information asynchronously with each other.

Because the system is distributed, we assume that 
each monitor only has access to a local clock, which
may slightly drift w.r.t~a~reference clock.
Hence the dates of the events are only known with a limited precision, and the exact order of the events occurring on different subsystems may be unknown.
We~assume that communications between local monitors are FIFO; this
way, when a monitor receives the information that some event took
place on some subsystem, it~also knows all the events that have
occurred earlier on that subsystem.

Thanks to these hypotheses, each monitor can determine a time in
the past for which it~currently has enough information to decide whether the
property it is checking already held, or failed to hold, at that time.
We consider real-time properties modelled as \emph{timed
automata}~(TA)~\cite{ALUR1994183}; the~expressiveness of these models
allows to account for the precise timing of events, which magnifies
the impact of the lack of an observable global clock.

\begin{example}
  \label{ex:ordering}
  \looseness=-1
  Consider a private (\eg, enterprise, university) network with a
    set of terminals and servers containing user account data logged
    in and a router connected to them. Properties of interest can be
    that (1)~all machines should remain connected to the network and
    (2)~no~account should be logged in more than one terminal
    at a~time.
    Even if two terminals signal that they are connected to a given account, a monitor has to wait to ensure that no "log out" message is pending from one of the two terminals. Plus, once such signals are received, the approximate timings may lead to situations where two terminals may or may not have been connected to the same user session at the same time, with no ability to conclude.%
\end{example}

\paragraph{Our contributions.}
We present 
a monitoring algorithm for properties expressed as (deterministic)~TAs in a distributed setting without a global clock.
Each~monitor keeps track of the most recent date 
at which it has collected the full history (relying on the assumption that
communications with other monitors are~FIFO). The~prefix of the
collected trace at this time already contains uncertainty, due
to the absence of a global clock, which entails that reordering
events from distinct components must be considered.
The computation of a verdict is based on the incremental update of a
structure that encodes the set of %
states
compatible with the prefix of the collected trace 
up to the time of interest.
We show that the monitoring algorithm is sound and complete.
\paragraph{Related works.}
\input{rw}

%% file: rw.tex
\label{sec:rw}

This paper 
studies distributed runtime verification for timed properties
described as timed automata~\cite{ALUR1994183}.
Several related approaches have been developed in the literature.
Extensions of Linear-time Temporal Logic~(LTL) integrating dense time have been explored for runtime verification and monitor synthesis~\cite{BauerLS11}.
Among~them, Metric Temporal Logic~(MTL) is of particular interest as it is directly related to 
TAs~\cite{NickovicM07} and is equipped with a progression function allowing to evaluate formulas at runtime~\cite{ThatiR05}.
Similarly, Timed Regular Expressions are as expressive as TAs and can be translated to TAs~\cite{AsarinCM02}. The~tool Montre~\cite{Ulus17} monitors them using timed pattern matching.
Monitoring TA models has been realized in the case of one-clock non-deterministic TAs~\cite{abs-2002-07049}.
Pinisetty~\emph{et~al.}~\cite{PinisettyJTFMP17} introduce a predictive setting for runtime verification of timed properties, leveraging reachability analysis to anticipate the detection of verdicts.

The aforementioned approaches consider that the monitored system is \emph{centralized} and the decision procedure is fed with a unique trace containing complete observations.
\emph{Decentralized} runtime verification~\cite{BauerF12} (see also a recent overview in~\cite{Falcone21})
handles separate traces corresponding to each monitor. 
Decentralized (also called synchronous) methods however assume the existence of a global clock shared by all components and monitors. 
We~relax this assumption here, and consider \emph{asynchronous distributed monitoring} (or~\emph{distributed monitoring} for~short).
For~a~discussion on the links and differences between synchronous and asynchronous methods, see~\cite{FrancalanzaPS18}.

Most approaches in decentralized runtime verification take as input Linear-time Temporal Logic formulas~\cite{BauerF12,ColomboF16,abs-2107-06084} or finite-state automata~\cite{FalconeCF14,El-HokayemF20}.
These approaches monitor specifications of discrete time, which does not account for the physical time that impacts the evaluation of the specifications nor the moment at which monitors perform their evaluation and deliver their verdicts.
An~approach close to ours is~\cite{Basin_2019}, in~which properties are specified in an extension of Metric Temporal Logic to tackle both timing and data values. Similar to~us, the~authors also deal with out-of-order messages, but also failures and lost messages.
However, they consider that local clocks are accurate.

Distributed runtime verification exhibits similarities with diagnosis~\cite{WangYL07,YinL15} (\cite{DBLP:journals/sttt/BouyerHJJM21} for~TAs), which aims to identify the occurrence of a fault and the component(s) responsible for it after a finite number of discrete steps, and has to cope with partial observation.
Our~approach differs from diagnosis, as we assume that monitors'(combined) local information suffices to detect violations; diagnosis does not aim to
check membership to an arbitrary timed regular expression.
The approach of~\cite{KohlHermanns23} for robust diagnosis shares several common aspects with~ours.
While centralized, it considers diagnosis where communication between the system and the diagnoser is subject to varying latency, clock drift
and out-of-order observation. 
The~problem is different but is similar in spirit: incrementally building a verdict based on approximate and partial timed observations.
Moreover some constructions have clear similarities, which is not surprising: in~both cases the language of runs compatible with the current partial and approximate observation has to be considered.
Other approaches comparable to ours perform monitoring on timed properties in a decentralized~\cite{DBLP:conf/time/RoussanalyF22} or 
distributed~\cite{GANGULY-JPDC-2024} fashion. 
Our own approach discusses distributed monitoring directly on timed automata models.

%% file: preliminaries.tex
\section{Preliminaries}
\label{sec:prelim}
We present the basic hypotheses about our formalization of the
distributed monitoring problem in \Cref{sub:pre_dec}, the notions of
words and languages at play in \Cref{sub:pre_tw,sub:pre_atw}, and 
timed automata in \Cref{sub:pre_ta}.

\subsection{Distributed timed systems}
\label{sub:pre_dec}

We consider systems  made of \(n\) independent
components~\((C_i)_{1\leq i\leq n}\). Each~$C_i$ is
observed by a local monitor~\(M_i\).  The~components being
independent, we~assume that each~$C_i$ has its own
finite alphabet of
actions \(\Act_i\), disjoint from the alphabets of the other components. 
We~write \(\Act=\bigsqcup_{1\leq i\leq n}\Act_i\) for the alphabet of all actions.
For~an action~$a \in \Act_i$, we~write $\Comp(a)=i$.
An~action~$a$ fired by component~$C_i$ is observed and
timestamped by its monitor~$M_i$,
giving rise to an \emph{event}~$(a,t)$ in $\Act_i \times \bbR+$.
We assume  the following  about the system:
\begin{description}
\item[Respective knowledge:] monitors know each others' alphabets of actions;

\item[Communication:] monitors send messages to their peers, carrying the timestamped events they observed (and events they learnt about from other monitors) in the order in which they observed them. We~assume that local events can be strictly ordered.

\item[Communication channels:] communications between monitors obey a FIFO policy with no message loss: messages are received in the order they were sent, and any sent message is eventually received, although no upper bound on communication delays is assumed.

\item[Communication topology:] the connectivity is such that a monitor~$M_i$ can receive messages (either directly or indirectly) from any monitor~$M_j$ managing some action appearing in~$M_i$'s property.

\item[Local liveness:] at any given time, each monitor will eventually have a local observation in the future, and will eventually send it to the other monitors.
\item[Time approximation:] monitors do not share a global clock, but one can assume that each local clock has a maximal skew $\sk$ with respect to a global \emph{reference clock}\footnote{Our approach can be easily generalized to different \(\sk\) for each monitor, with our theorems depending on the greatest one.}. We~suppose that clocks are non-decreasing.
\end{description}

Most of these assumptions are easy to satisfy. A~FIFO policy can be
achieved by numbering the events exchanged between one monitor and
another, ensuring that a monitor can handle events in the order they
were sent. Local liveness can be ensured by adding empty events that are
sent when no events have been observed for a long duration. The~main practical constraints are the absence of message losses, and the bound on clock skews.
However, even these assumptions can be mitigated. Message losses can be detected using the message numbering, while  a mechanism such as Network Time Protocol~(NTP) can be used to limit the skew to~\(\sk\).

We~consider
that each monitor is in
charge of verifying some property (given as a timed automaton, see \cref{sub:pre_ta}). 
As the monitoring algorithms and properties are symmetric for each such monitor,
we~restrict our dissertation to a fixed monitor~$M_i$ and assume that the entire set \(\Act\) of actions is necessary for its property (some of these actions are still observed by other monitors).

\subsection{Timed words and languages}
\label{sub:pre_tw}

We consider intervals in $\bbR$, denoted $\Interv(\bbR$).
For an interval $I=\langle l,u \rangle$, with~${\mathord\langle \in \{\mathord(,\mathord[\}}$
and  $\mathord\rangle \in \{\mathord),\mathord]\}$, 
we~write $\lb(I)=l$, $\ub(I)=u$ for its lower and upper bounds.   
For~two intervals~$I_1, I_2$ of~$\bbR$, we~write~${I_1 \prec  I_2}$
if $\ub(I_1)\leq\lb(I_2)$; when this condition is met,
the intervals intersect in at most one point.
We also consider intervals of natural numbers and write $\IntInv{m}$ for $\{k \in \mathbb{N} \mid 1\leq k \leq m\}$.

\paragraph{Timed Words and Languages.}\looseness=-1
We consider \emph{timed words} built on the alphabet of actions~$\Act$
as (finite or infinite) sequences
\(\tw=(a_k,t_k)_{k\in \IntInv{m}}\)\footnote{We abusively use such a notation for both finite and infinite sequences.}
of \emph{events} in~\(\Act\times\bbR+\) 
whose sequence of \emph{dates} $(t_k)_{k\in \IntInv{m}}$  is non-decreasing.
We~write~\(\tw[k]=(a_k,t_k)\) for its $k$-th event.
For~any interval~$I$ of~$\bbR+$, we~write~$\tw_{\mid I}$ for the subword of~$\tw$ restricted to dates in~$I$,
and~\(\tw_{\mid t}\) as a shorthand for~\(\tw_{\mid [0,t]}\) with~$t\in\bbR+$.
For~a~finite timed word \(\tw=(a_k,t_k)_{k\in \IntInv{m}}\), we~write \(|\tw|=m\) for its length, $\firstt(\tw)=t_1$ and $\lastt(\tw)=t_m$ respectively for the dates of its first and last events;
for the empty word~$\emptyw$, we~let $\firstt(\emptyw)=\lastt(\emptyw)=0$.
Two~timed words $\tw_1$ and $\tw_2$ can be concatenated into $\tw_1\cdot\tw_2$ if, and only~if, $\lastt(\tw_1)\leq \firstt(\tw_2)$.
For any interval~$I\subseteq \Rplus$, we~write \(\TT_I(\Act)\) for the
set of timed words in \((\Act\times I)^{*}\), and \(\TT(\Act)=\TT_{\Rplus}(\Act)\).
A~language of timed words is any subset of $\TT(\Act)$.

For two languages of timed words $L_1$ and $L_2$, their concatenation $L_1\cdot L_2$ is defined and equal to
$\{\tw_1\cdot \tw_2 \mid \tw_1 \in L_1, \tw_2\in L_2\}$
if, and only~if,
$\supp\{\lastt(\tw_1)\mid\tw_1 \in L_1\} \leq \inf\{\firstt(\tw_2)\mid\tw_2 \in L_2\}$.
The~restriction to an interval naturally generalizes to languages:
for a language $L\subseteq \TT(\Act)$ and an interval $I$ of $\bbR$,   \(L_{\mid I}=\{\tw_{\mid I} \mid {\tw \in L}\}\).

\paragraph{Projections on monitors.}

In our setting, $\Act$~is the disjoint union of alphabets~$\Act_i$.
For~a timed word~$\tw$, we~write~\(\proj_i(\tw)\) for the projection
on the actions monitored by~\(M_i\), defined by induction as
\(\proj_i(\emptyw)=\emptyw\) for the empty word, and
\(\proj_i(\tw\cdot(a,t))=\proj_i(\tw)\cdot(a,t)\) if~\(a\in\Act_i\)
and \(\proj_i(\tw\cdot (a,t))=\proj_i(\tw)\) otherwise.

Conversely, we define a tensor operation on timed words \(\tw_1\tens\tw_2\) that merges the events while re-ordering them by ascending timestamps.
This operation is such that \({\tw=\mathop\tens_{i \in \IntInv n}\proj_i(\tw)}\).

\subsection{Approximate Timed Words}
\label{sub:pre_atw}

If we had perfect clocks, timed words as defined above would be the
model of choice for representing the knowledge of the monitors;
restriction to intervals would be used to identify the part of the
knowledge that each monitor knows is complete (as opposed to the part
of the knowledge where informations from some of the components did
not arrive~yet).

\looseness=-1
In the context of distributed monitoring, we~consider that the clocks
of the monitors may be imprecise, resulting in a potential drift of
up~to a uniform bound~$\sk$ w.r.t.~a~reference clock. Because of this,
we~have to rely on a notion of \emph{approximate timed words}, and to define
restriction to intervals for this new model.

Because of timing imprecisions, an~event~$(a,t)$,
made of action~$a$ timestamped with~$t$ by the monitor that observed~it,
may~have happened anywhere in the interval $[t-\sk,t+\sk]\cap\bbR+$ with respect to the reference clock\footnote{Intersection with $\bbR+$ is used to rule
out events with negative dates.}.
Thus, while the information collected by a monitor has the form of
a timed word \(\tw=(a_k,t_k)_{k\in\IntInv{m}}\),
the~real date of each event $(a_k,t_k)$ lies in the interval
$I_k=[t_k -\sk, t_k +\sk]\cap\bbR+$.
We~call \emph{approximate timed word~of}~$\tw$,
the sequence \(\atw(\tw)={(a_k,I_k)}_{k\in\IntInv{m}}\).

\paragraph{Approximate timed words.}
An~\emph{approximate timed word} (ATW~for short) is a sequence of pairs $\atw=(a_k,I_k)_{k\in\IntInv{m}}$  such that, for all~$k\in\IntInv{m}$, $a_k \in \Act$ and $I_k\in \Interv(\bbR)$~is an interval (open or closed for generality).
We~denote by $\ATW(\Act)$ the set of approximate timed words on~$\Act$.

With any~approximate timed word, we~associate two languages:
its~\emph{ordered} language and its \emph{non-ordered} language.
Intuitively, the~\emph{ordered language} of~\(\atw\) is the language
of timed words that respect the order of the events and the intervals
given by~$\atw$, while the \emph{non-ordered language} will be the
union of the ordered languages for all possible reoderings of the
events.

The ordered semantics defines a ``tube'' of timed words with the same
untimed projection. It~is defined as follows:
\begin{equation*}
\ordlang{(a_k,I_k)_{k\in\IntInv{m}}} = 
\bigl\{ 
  (a_k,t_k)_{k\in\IntInv{m}}\in \TT(\Act) 
  \mid \forall k\in \IntInv{m}.\ t_k \in I_{k}
\bigr\} 
\end{equation*}
By definition of $\TT(Act)$, the~sequence~$(t_k)_{k\in\IntInv{m}}$ is
non-decreasing, which induces constraints on subsequent $t_k$'s. 
Notice that $\ordlang{\nu}=\emptyset$ when $I_k=\emptyset$ for some~$k$,
or if no increasing sequences of dates can
be found in the sequence~$(I_k)_{k\in \IntInv{m}}$.

In order to define the \emph{non-ordered language} of~$\atw$, we~introduce
the~subset~$\mathcal{F}(\atw)$ of permutations of events
in~$\atw$ that respect the strict order of events sharing the same component
(each monitor knows the order of events occurring in the component it supervises).
We~define~it as~follows, with \(\Perms(\IntInv{m})\) being the set of permutations of \(\IntInv{m}\):
\begin{xalignat*}1
\mathcal{F}((a_k,I_k)_{k\in \IntInv{m}}) =  \bigl\{ & 
\perm\in\Perms(\IntInv{m})
\mid
\forall k,l\in \IntInv m. \\&
 (\Comp(a_k)=\Comp(a_l) \wedge  k<l) \Rightarrow f(k)<f(l)
\bigr\}.
\end{xalignat*}
Then for $f\in\mathcal{F}(\atw)$,
we~abuse the notation and write $f(\atw)$ for the ATW $(a_{f(k)},I_{f(k)})_{k\in \IntInv{m}}$.
Finally, we~define the \emph{(non-ordered) language} of~$\atw$
as the set of timed words that respect both the intervals given by~$\atw$,
and the strict local order on each component:
\[
\nolang{\atw} =  \bigcup_{f \in \mathcal{F}(\atw)}\ordlang{f(\atw)}.
\]

Intuitively, 
this language includes commutations of events that occurred at sufficiently close dates on different components: 
indeed, if
\(\atw=\atw_1\cdot (b_k,t)\cdot (b_{l},t')\cdot\atw_2\) with $\Comp(b_k)\neq \Comp(b_{l})$ and $|t-t'|\leq 2\sk$, then 
\(\nolang{\atw}\) contains timed words with form \(\tw_1\cdot (b_k,t)\cdot (b_{l},t')\cdot\tw_2\) 
and \(\tw_1\cdot (b_{l},t'-\sk)\cdot (b_{k},t+\sk)\cdot\tw_2\). 

Back to the monitoring problem, given some monitor and an observation prefix~$\tw$ that is sufficient (in some sense clarified later),
considering the imprecision due to the skew $\sk$ in the approximate timed word $\atw(\tw)$, 
the non-ordered language $\nolang{\atw(\sigma)}$
is the set of executions that could produce this observation prefix, hence that has to be considered for monitoring.

\paragraph{Operations on approximate timed words.}
We~now focus on defining a restriction of approximate timed words to an
interval of dates.
This will be useful for the incremental update of the monitor's knowledge (see \cref{sub:CS}).
Semantically, the~restriction of an ATW~$\atw$ to an interval~$I$ is the set of restrictions to~$I$ of all
the timed words contained in~$\nolang{\atw}$.
In~this section, we~present a syntactic
definition, which will be the basis of an effective computation.

To~this aim, we~first define \emph{intersection}: 
for~\(\atw= (a_k,I_k)_{k\in \IntInv{m}}\) and an interval~$I$,
the~intersection of~$\atw$ with~$I$ is the ATW \(\atw_{\cap
  I}=(a_k,I_k\cap I)_{k\in \IntInv{m}}\).
Notice that $\ordlang{\nu_{\cap I}} = \nolang{\atw_{\cap I}} = \emptyset$ if $I_k\cap I=\emptyset$ for some~$k\in\IntInv{m}$.

Our syntactic definition of restriction adapts the notion of subword.
For~any interval~$I$ of~$\bbR+$,
and two \ATW~$\atw'$ and~$\atw''$, we say that
\(\atw'=\atw'_1 \cdots \atw'_n\) is a \emph{subword of} 
\(\atw''=\atw''_1\cdot\atw'_1 \cdots \atw''_n\cdot\atw_n'\cdot\atw''_{n+1}\) \emph{conditioned by~$I$},
written $\atw' \preceq_I \atw''$,  if, and only if,
\begin{itemize}
\item for all \(l \in  \IntInv{n+1}\), 
  for any \((a_k,I_k)\) in \(\atw''_l\), \(\neg(I_k \subseteq I)\):
  all events in~$\atw''_l$ (which are dropped) \emph{may occur outside of~$I$};
\item for all \(l \in \IntInv{n}\),
  for any \((a'_{k'},I'_{k'})\) in \(\atw'_l\),  \(I'_{k'}\cap I\neq\emptyset\):
  all events in~$\atw'_l$ (which are not dropped) \emph{may occur in~$I$};
\item there is \(f\in\calF(\atw'')\) %
  s.t. \(f(\atw'')=\atw_1\cdot\atw'\cdot\atw_3\) for some~$\atw_1$ and~$\atw_3$: 
  this encodes the fact that two events in the same component can not be permuted. 
\end{itemize}

We can now define the (syntactic) \emph{restriction} of an
approximate timed word~$\atw$ to an interval~$I$ as $\atw_{\mid I}
=\{\atw'_{\cap I} \mid \atw' \preceq_I \atw\}$. As a shorthand, for a timestamp $T$, we write $\atw_{\mid [0,T]} =\atw_{\mid T}$. We~overload the term
\emph{restriction} of~\(\atw\) to~\(I\) because of the
characterization of Lemma~\ref{lm:atw_restriction} below:
the~syntactic restriction corresponds to
the semantic approach of taking the language of timed words
associated with~\(\atw\), and restricting each of its words.
This~provides us with a way of representing, manipulating and
computing restriction of ATW to intervals:
\begin{restatable}{lemma}{atwrestriction}
  \label{lm:atw_restriction}
  For any approximate timed word \(\atw\) and any timestamp~$T$,
  it~holds
  \(\cup_{\atw'\in\atw_{\mid T}} \nolang{\atw'} = %
                \nolang{\atw}_{\mid T}\).
\end{restatable}
Following this, we write \(\nolang{\atw_{\mid T}}\) for \(\cup_{\atw'\in\atw_{\mid T}} \nolang{\atw'}\).

\begin{example}
  \label{ex:atw_op}
  Consider
  the approximate timed word $\atw=(a,[1,3])(b,[2,4])(c,[3,5])$, and $I=[0,3]$ with the components of the actions being pair-wise different. Then:%
  \begin{itemize}
  \item $\atw_{\cap I}$ is the approximate timed word $(a,[1,3])(b,[2,3])(c,\{3\})$. 
  \item assuming that all three events occur on different components
    (and can then be freely swapped), the~set $\{\atw' \mid
    \atw'\preceq_I \atw\}$ is
    \begin{multline*}
    \{\epsilon, (a,[1,3]), (a,[1,3])(b,[2,4]), (a,[1,3])(c,[3,5]), \\
    (a,[1,3])(b,[2,4])(c,[3,5]), (b,[2,4]), (b,[2,4])(c,[3,5]),
    (c,[3,5])\}
    \end{multline*}
    (in~other terms, all subwords are allowed, since all intervals
    intersect~$I$ and none of them are included in~$I$).
    It~follows that   $\atw_{|I}$ is the union of
    $\epsilon$, $(a,[1,3])$, $(a,[1,3])(b,[2,3])$,
    $(a,[1,3])(c,[3,3])$, $(a,[1,3])(b,[2,3])(c,[3,3])$, $(b,[2,3])$,
    $(b,[2,3])(c,[3,3])$ and $(c,[3,3])$.

    Now, assume that~$b$ and~$c$ relate to the same component, so
    that they cannot be swapped.
    In~this case, $(a,[1,3])(c,[3,5])$ is no longer a subword of~$\atw$ for~$I$,
    because the third condition is no longer fulfilled.
  \end{itemize}
\end{example}

\subsection{Formalism for timed systems}
\label{sub:pre_ta}
\looseness=-1
We monitor properties given as TAs, which
are standard formalism for expressing properties of time-constrained
systems: the~aim of our monitoring procedure is to decide if the execution
we are (partially and imprecisely) observing is (or~will~be) accepted by a given
timed automaton. We~introduce the formalism of timed automata in thie section.

Let~\(\Clocks\) be a finite set of clocks.  A~clock
valuation is a function \(\val\colon
\Clocks\rightarrow\Rplus\).
We~write $\Vals(\Clocks)$ for the set of valuations. 
The~initial valuation is \(\initv\colon \clock\in\Clocks\mapsto 0\); a~time
elapse for a delay \(\te\in\Rplus\) maps valuation~$\val$ to
\(\val+\te\colon\clock\mapsto\val(\clock)+\te\), and a clock reset for
a subset \(\Clocks'\subseteq\Clocks\) maps $\val$ to
\(\val_{[\Clocks']}\) such that
\(\val_{[\Clocks']}(\clock)=\val(\clock)\) if $\clock\notin\Clocks'$,
and $\val_{[\Clocks']}(\clock)=0$ otherwise.

A \emph{zone} is a finite conjunction of clock constraints of the
forms \(\clock_1\bowtie n\) and \(\clock_1-\clock_2\bowtie n\), with
\(\clock_1,\clock_2\in\Clocks\), \(\mathord{\bowtie}
\in\{\mathord<,\mathord\leq,\mathord=,\mathord\geq,\mathord>\}\) and
\(n\in\bbN\). We~write \(\Zones(\Clocks)\) for the set of zones
of~\Clocks. We~write \(\val\models\zone\) and say that \val
\emph{satisfies} \zone when the values of the clocks in \val satisfy
the constraints in \zone.

\begin{definition}
A \emph{timed automaton} (TA) is a tuple \(\ta=(\Loc,\initloc,\Act,\Clocks,\Trans)\) where \Loc is a finite set of \emph{locations} containing  the initial location~\initloc, \Act and \Clocks are finite sets of actions and clocks respectively, and  \(\Trans\subseteq\Loc\times\Zones(\Clocks)\times\Act\times2^{\Clocks}\times\Loc\) is the set of\emph{transitions}. 
For a transition \((\loc,\guard,a,z,\loc')\in\Trans\), we call \loc its \emph{source}, \(\loc'\) its \emph{target}, \(\guard\) and $a$ its \emph{guard} and \emph{action} and \(z\) its \emph{reset~set}.
We~call \emph{configurations of~$\ta$} the triples \((\loc,\val,t)\in\Loc\times\Vals\times\bbR+\), and \emph{initial configuration} the configuration~\((\initloc,\initv,0)\). 
\end{definition}
\begin{remark}
  We add to the usual notion of configuration a date representing the
  instant at which the system is in this configuration (after a given
  behaviour). We~do~this because our reasoning is based on timestamps, not delays, following the definition of timed words based on observations. This can be readily implemented by giving an additional clock (which is never reset) to the~TA.  
\end{remark}

A timed word \(\tw=(a_i,t_i)_{1\leq i}\) is \emph{a trace} of a timed automaton if,
starting from the initial configuration \((\initloc,\initv,0)\), one can find
a sequence \(\run=((\loc_{i-1},\val_{i-1},t_{i-1})\xrightarrow{\delta_i}(\loc_{i-1},\val_{i-1}+\delta_i,t_{i-1}+\delta_i)\xrightarrow{\trans_i}(\loc_i,\val_i,t_{i-1}+\delta_i))_{1\leq i}\) with 
\(\trans_i=(\loc_{i-1},\guard_i,a_i,z_i,\loc_i)\in\Trans\), 
 \(\delta_i = t_i-t_{i-1}\) (with \(t_0=0\)),
 and at each step \(\val_{i-1}+\delta_i\models \guard_i\)  and \(\val_{i}=(\val_{i-1}+\delta_i)_{[z_i]}\).
 Such a sequence $\run$ is called a (finite or infinite) \emph{run}, and we write \(\trace(\run)=\tw\).
 A~timed word \tw~is a \emph{partial} trace if there exists a timed word~\(\tw'\) such that \(\tw'\cdot\tw\) is a trace.
 A partial trace thus corresponds to a \emph{partial run} that does not (necessarily) start from \((\initloc,\initv,0)\).
 For a (partial) trace~\tw with $\firstt(\sigma)\geq t$, we write \((\loc,\val,t)\stackrqarrow{\tw}{t}(\loc',\val',t')\) when \tw leads from configuration~$(\loc,\val,t)$ to configuration \((\loc',\val',t')\).

A~timed automaton is said \emph{deterministic} 
if for any two transitions $(\loc,g,a,z,\loc')$ and
$(\loc,g',a,z',\loc')$ such that $g\cap g'\not=\emptyset$, it~holds
$g=g'$ and~$z=z'$. It~is said \emph{complete} when for any
configuration \((\loc,\val,t)\) and action \(a\in\Act\), there is at
least a transition \((\loc,g,a,z,\loc')\) such that \(\val\models g\).
When considering complete deterministic automata, the~\trace function
defined above is a bijection, and we identify traces with their
associated runs.
In this context, we~add the possibility of having a final delay after a trace:
for a configuration $(\loc,\val,t)$, a date $t''$ and a (partial) trace $\tw$  such that \(\firstt(\tw)\geq t\) and \(\lastt(\tw)=t'\leq t''\) we write  $(\loc,\val,t) \;\after_t^{t''}\; \tw$ for the unique configuration $(\loc',\val'',t'')$ such that 
\((\loc,\val,t) \stackrqarrow{\tw}{t}(\loc',\val',t')\xrightarrow{t''-t'}(\loc',\val'',t'')\), \ie, reached from \((\loc,\val,t)\) 
after the trace $\tw$ followed by the delay $t''-t'$.
This can be generalized to sets of configurations and languages: 
for a set of configurations~$S$ and a language of timed words~$L$, 
such that $\inf\{\firstt(\tw)\mid \tw \in L\}\geq t$ and \(\supp\{\lastt(\tw)\mid \tw \in L\}\leq t''\), 
 \(S \;\after_t^{t''}\; L =\{(\loc',\val',t'') \mid \exists (\loc,\val,t)\in S, \exists \tw \in L,  (\loc',\val',t'')= (\loc,\val,t) \; \after_t^{t''}\; \tw\}\).
  Finally, we~define the corresponding notion when no date is given: 
  \(
  S \;\after\; L =\{(\loc',\val',t') \mid \exists (\loc,\val,t)\in S, \exists \tw \in L, \exists t', (\loc',\val',t')= (\loc,\val,t) \; \after_t^{t'}\; \tw\}
\).
  Notice that in this last definition, $t$ and $t'$ are not bound and can range on all values such that \(\after_t^{t'}\) is defined.

\paragraph{Modelling properties.}
The~property associated with monitor~$M_i$
is defined by a deterministic and complete~TA $\ta_i$ %
and a subset \(\Final_i\) of locations specifying a reachability property\footnote{We here restrict to deterministic and complete TAs for simplicity, but generalization to non-deterministic and incomplete TAs is easy.}:
we~write~$\sem{\ta_i,\Final_i}$ for the set of (runs~of) finite traces~$\tw$ that end in some location of~$\Final_i$ when applied to $\ta_i$ from its initial configuration~\((\initloc,\initv,0)\);
we~extend~$\sem{\ta_i,\Final_i}$ (abusively keeping the same notation) to include runs of infinite traces~$\tw$ for which
there is a length~$k$ such that all prefixes of~$\tw$ of length larger than~$k$ are in~$\sem{\ta_i,\Final_i}$. 

Given a property specified by $\ta_i$ and $\Final_i$, a finite trace \tw is a \emph{good prefix} (resp. \emph{bad prefix}) if for all \emph{infinite}
continuations \(\tw\cdot\tw'\in(\Act\times\Rplus)^\omega\) of \tw, \(\tw\cdot\tw'\in\sem{\ta_i,\Final_i}\) (resp. \(\tw\cdot\tw'\notin\sem{\ta_i,\Final_i}\)).
In~terms of automata, this means that the prefix reached some configuration in~$\Loc_i\times \Vals\times\bbR+$ from which it will always eventually stay
in~$\Final_i \times \Vals\times\bbR+$ (resp.~it~never visited and will never visit~$\Final_i$).
We note this set of configurations \(\Inev(\Final_i)\) (resp. \(\Never(\Final_i)\)).
Good prefixes (resp. bad prefixes) are then traces of runs in~$\sem{\ta_i,\Inev(\Final_i)}$ (resp. $\sem{\ta_i,\Never(\Final_i)}$). 
Starting from~$\Final_i$, the state sets \(\Inev(\Final_i)\) and  \(\Never(\Final_i)\)
can be computed off-line by a zone-based co-reachability analysis~\cite{BY04}.
Thanks to this, we~restrict our focus to the reachability of pairs of locations and zones without loss of generality.
\looseness=-1
These notions can be extended to languages, thus to approximate timed words.
A language $L\in \TT_{\Act}$ is a \emph{good} (resp.~\emph{bad}) \emph{language prefix} if %
$L \subseteq \sem{\ta_i,\Inev(\Final_i)}$ (resp.  $L \subseteq \sem{\ta_i,\Never(\Final_i)}$).
These can also be computed using \(\Inev(\Final_i)\) and  \(\Never(\Final_i)\).

%% file: base.tex
\section{Monitoring with complete information}
\label{sec:tmin}
The role of monitoring algorithms is to provide us with verdicts when analyzing executions of the system.
Since we want this to be performed online, verdicts should be given as soon as possible, based on the observation of a finite execution prefix.
However, in~the~context of distributed systems, the~observation collected by a monitor at a given date may be imperfect, with missing events and approximate dates.
We~first identify the points in time where we have enough information to decide a verdict in~\cref{sub:tmin_def}, and then define our verdicts of interest in \cref{sub:verdict_tmin}. Using~this, we~explain the data structure we~use and its related operations in~\cref{sub:CS}, and explain how to compute a verdict
in \cref{sub:monit_tmin}.%

\subsection{Point of Certainty}
\label{sub:tmin_def}

When the components of the system perform actions, their corresponding monitors 
\((M_j)_{j \in \IntInv{n}}\) instantly \emph{observe} these actions and timestamp them with the value of their (local) clock. However, they need to wait for the communication of other monitors in order to \emph{collect} the information about the other components' events.

\paragraph{Communication policy.}
We consider the simple policy in which each monitor~$M_j$ instantly sends its observations (action performed and timestamp) to every other monitor $M_i$ that needs~it for checking its property, grouping in the same message all the events that occurred at the same instant\footnote{This technical detail is useful for \cref{prop-tmin1} to ensure that all events of same date issued from the same component are collected simultaneously.
It can be implemented by waiting any non null delay before sending a message aggregating the observations.}.
As there are no bounds in communication delays, monitors still have to deal with partial and out-of-order information.
Moreover, local time approximation induces imprecision in event dates.
To make monitoring sound, we first determine the time point at which we can safely monitor with no missing event. 

Formally, consider the \emph{global observation} \(\two\) such that at a given global time \(t^g\), the observation collected by all monitors is hence \(\two_{\mid t^g}\). 
We know that the \emph{global trace} \(\twg\) of the run of the system is such that \(\twg\in\nolang{\atw(\two)}\). 
This trace can not be observed, yet it is the one we want to monitor, hence we will start our reasoning from (prefixes of) the language \(\nolang{\atw(\two)}\).

 Moreover, no single monitor has access to \(\two_{\mid t^g}\) %
 at time $t^g$ due to the need for synchronization.
 Let the \emph{collected trace} at $t$ by $M_i$, written \(\tw_i(t)\), be the monitoring information gathered by a monitor $M_i$ at the $M_i$-local time~$t$.
 It is composed of a subset of the global observation \(\two_{\mid t+\sk}\), containing 
 at least its own local observation \(\proj_i(\two)_{\mid t}\) and events received from other monitors, forming for each monitor $M_j$ a timed word \(\proj_j(\two)_{\mid t_j}\) with \(t_j\leq t+2\sk\).
 Indeed, communication being FIFO, $M_i$ receives the information from each individual $M_j$ in order, but potentially with $M_j$-local timestamps up to $t+2\sk$, as both the $M_i$-local time $t$ and the $M_j$-local time can skew by a maximum of $\sk$ from the global time.

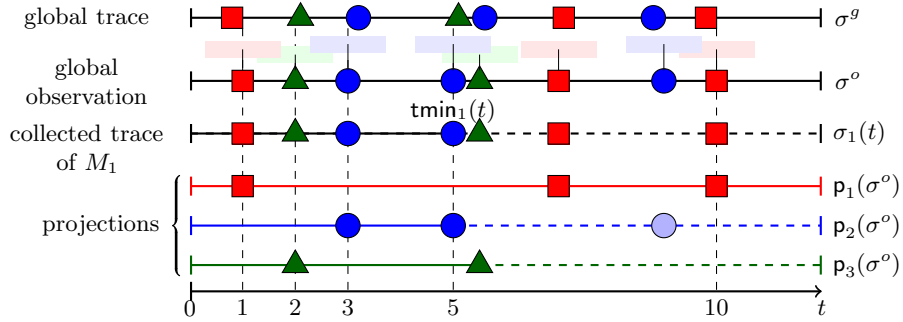
\begin{figure}[t]
\centering
\begin{tikzpicture}[scale=.7, %
triang/.style = {regular polygon, regular polygon sides=3,
              draw, fill=green!40!black, text width=0em, 
               minimum size = 0.4cm, inner sep=0pt, outer sep=0mm,
              shape border rotate=0},
sqare/.style = {regular polygon, regular polygon sides=4,
              draw, fill=red, text width=0em, 
               minimum size = 0.4cm, inner sep=0pt, outer sep=0mm,
              shape border rotate=0},
cir/.style = {circle,
              draw, fill=blue, text width=1em,
               minimum size = 0.1cm, inner sep=0pt, outer sep=0mm,
               shape border rotate=0},
rect/.style = {regular polygon, regular polygon sides=4,
              draw, fill=grey, text width=0em, 
               minimum size = 0.4cm, inner sep=0pt, outer sep=0mm,
              shape border rotate=0}
  ]

 \draw [thick,|->](0,-0.5) node[black,below] {$0$ } -- (12,-0.5) node [black, below] {$t$};
 \draw [dashed] (1,4.5) -- (1,-0.5) node[below] {$1$};
 \draw [dashed] (2,4.5) -- (2,-0.5) node[below] {$2$};
 \draw [dashed] (3,4.5) -- (3,-0.5) node[below] {$3$};
 \draw [dashed] (5,4.5) -- (5,-0.5) node[below] {$5$};
 \draw [dashed] (10,4.5) -- (10,-0.5) node[below] {$10$};
\draw [thick,|-,green!40!black](0,0) node[black,below=1mm] {$ $} -- (5.5,0) node [black, xshift=0cm, yshift=.3cm] {};
\draw [thick,dashed,-|,green!40!black](5.5,0) -- (12,0) node [black, xshift=.6cm, yshift=0cm] {$\proj_3(\tw^o)$} node[black,below=1mm] {$ $};
\node [triang] (t1) at (5.5,0) {};
\node [triang] (t2) at (2,0) {};

\draw [thick,blue,|-](0,.75) -- (5,.75) node [black, xshift=0cm, yshift=.3cm] {};
\draw [thick,dashed,-|,blue](5,.75) -- (12,.75) node [black, xshift=.6cm, yshift=0cm] {$\proj_2(\tw^o)$};
\node [cir] (c1) at (3,.75) {};
\node [cir,fill =blue!30] (c2) at (9,.75) {};
\node [cir] (c3) at (5,.75) {};

\draw [thick,|-|,red](0,1.5) -- (12,1.5) node [black, xshift=.6cm, yshift=0cm] {$\proj_1(\tw^o)$};
\node [sqare] (s1) at (1,1.5) {};
\node [sqare] (s2) at (7,1.5) {};
\node [sqare] (s3) at (10,1.5) {};

\begin{scope}%
  \draw [thick,|-](0,2.5) node[left,text width=1.5cm,align=center] { } -- (5,2.5) node [black, xshift=0cm, yshift=.3cm] {$\tmin_1(t)$};
  \draw [thick,dashed, -|](0,2.5) node[left,text width=2.5 cm,align=center,yshift=-.2cm] {collected trace  \\ of $M_1$} -- (12,2.5) node [black, xshift=.5cm, yshift=0cm] {$\tw_1(t)$};%
  \node [triang] (t1b) at (5.5,2.5) {};
  \node [triang] (t2b) at (2,2.5) {};
  \node [sqare] (s1b) at (1,2.5) {};
  \node [sqare] (s2b) at (7,2.5) {};
  \node [sqare] (s3b) at (10,2.5) {};
  \node [cir] (c1b) at (3,2.5) {};
  \node [cir] (c3b) at (5,2.5) {};
\end{scope}

\draw [thick, |-| ](0,3.5) node[left,text width=2.5cm,align=center] {global \\ observation} -- (12,3.5) node [black, xshift=.35cm, yshift=0cm] {$\tw^o$};%
\node [triang] (t1b) at (5.5,3.5) {};
\node [triang] (t2b) at (2,3.5) {};
\node [sqare] (s1b) at (1,3.5) {};
\node [sqare] (s2b) at (7,3.5) {};
\node [sqare] (s3b) at (10,3.5) {};
\node [cir] (c1b) at (3,3.5) {};
\node [cir] (c2b) at (9,3.5) {};
\node [cir] (c3b) at (5,3.5) {};

\node[rectangle,
  minimum width = 1cm,
    minimum height = .1cm,
	fill = green!10] (r) at (5.5,4) {};
\node[rectangle,
  minimum width = 1cm,
  minimum height = .1cm,
	fill = green!10] (r) at (2,4) {};
\node[rectangle,
  minimum width = 1cm,
    minimum height = .1cm,
    fill = red!10] (r) at (1,4.1) {};
\node[rectangle,
  minimum width = 1cm,
    minimum height = .1cm,
	fill = red!10] (r) at (7,4.1) {};
\node[rectangle,
  minimum width = 1cm,
    minimum height = .1cm,
	fill = red!10] (r) at (10,4.1) {};
\node[rectangle,
  minimum width = 1cm,
    minimum height = .1cm,
    fill = blue!10] (r) at (9,4.2) {};
\node[rectangle,
  minimum width = 1cm,
    minimum height = .1cm,
    fill = blue!10] (r) at (3,4.2) {};
\node[rectangle,
  minimum width = 1cm,
    minimum height = .1cm,
    fill = blue!10] (r) at (5,4.2) {};

\draw[line width=.2pt] (5.5,3.7) -- (5.5,4);
\draw[line width=.2pt] (2,3.7) -- (2,4);
\draw[line width=.2pt] (1,3.7) -- (1,4.1);
\draw[line width=.2pt] (7,3.7) -- (7,4.1);
\draw[line width=.2pt] (10,3.7) -- (10,4.1);
\draw[line width=.2pt] (9,3.7) -- (9,4.2);
\draw[line width=.2pt] (3,3.7) -- (3,4.2);
\draw[line width=.2pt] (5,3.7) -- (5,4.2);

\draw [thick, |-| ](0,4.7) node[left,text width=2.5cm,align=center] {global trace} -- (12,4.7) node [black, xshift=.35cm, yshift=0cm] {$\tw^g$}  node[black,below=1mm ] {$ $};
\node [triang] (t1b) at (5.1,4.7) {};
\node [triang] (t2b) at (2.1,4.7) {};
\node [sqare] (s1b) at (.8,4.7) {};
\node [sqare] (s2b) at (7.1,4.7) {};
\node [sqare] (s3b) at (9.8,4.7) {};
\node [cir] (c1b) at (3.2,4.7) {};
\node [cir] (c2b) at (8.8,4.7) {};
\node [cir] (c3b) at (5.6,4.7) {};

\draw[line width=1pt,decorate, decoration = {calligraphic brace}] (-.2,-.2) -- (-.2,1.7)
  node[midway,left,text width=1.7cm] {projections};
\end{tikzpicture}
\caption{A global finite trace~$\tw^g$ (top), its corresponding global observation $\tw^o$ at local $M_i$-time $t$ (below) with rectangles figuring global time approximation ($\sk=0.7$), 
the collected trace $\tw_1(t)=({\color{red} a}, 1)({\color{green!40!black} c}, 2)({\color{blue} b}, 3)({\color{blue} b}, 5)({\color{green!40!black} c}, 5.5)({\color{red} a}, 7)({\color{red} a}, 10)$  (middle), \ie,  observation of $M_1$ completed with some events received from $M_2$ and $M_3$,
  the projections observed locally by three monitors $M_1$, $M_2$, $M_3$ at~$t$ (bottom).
  Dashed lines represent information uncertainty, \eg, $M_1$ ignores what happened after the last event received from each of the other monitors.
}
\label{fig-extmin}
\end{figure}

For $M_i$ at $M_i$-local date $t$, consider the set 
\[\{(j,t_j)  \in \IntInv{n} \times \Rplus \mid j\neq i \wedge \left(\lastt(\proj_j(\tw_i(t)))=t_j\right)\}\cup\{(i,t)\}\]
of pairs made of the index $j$ of each monitor coupled with the timestamp of the last observation received from~$M_j$ by~$M_i$.
Let \((\jtmin_k)_{k\in\IntInv{n}}=(\jmin_k,\tmin_k)_{k\in \IntInv{n}}\) be the sequence 
obtained by ordering this set of pairs by ascending timestamp~\footnote{We should write
  \((\jtmin_k(t))_{k\in \IntInv{n}}=(\jmin_k(t),\tmin_k(t))_{k\in \IntInv{n}}\), \ie, parametrize  by~$t$,
  but we will often forget~$t$ when clear from the context.}.
Initially, those timestamps are all~$0$, and any order may be chosen.
Then, this sequence with its ordering can be easily maintained on-the-fly when new events are observed or received by~$M_i$.
Clearly, $(\jmin_1,\tmin_1)$ identifies the monitor~$M_j$ for which
$M_i$~is aware of the earliest timestamp (in~its local time).
In~the absence of a~skew, $M_i$~would be sure to have complete information from all other monitors at time~$\tmin_1$, with $\jmin_1$ being the monitor for which the last event known by~$M_i$ (if~any) is the oldest.
However, as time is approximated, and since verdicts should be given on global traces, knowing all events at local times up to $\tmin_1$ only certifies that all events have been recorded for global time up to \(\tmin_1-\sk\), as seen in \cref{ex:tmin_sk}.
Notice that this also entails that the verdict can be given at $t$ only if $t \geq \sk$.
\begin{example}
  \label{ex:tmin_sk}
  In~\cref{fig-extmin}, for monitor $M_1$, $\jmin_1=2$ at time~$t$. Yet, if the verdict was given with respect to observations after the last event from~$M_2$,
  the~last event from~$M_3$,  which happened before~it but was marked with a later timestamp, would be missed. 
  The~verdict should thus restrict to the earliest possible global date for the last event of $M_2$, namely $\tmin_1-\sk$.
\end{example}

Conversely, we are sure that for all monitors, we collected some event with timestamp at least $\tmin_1$.
As local times are non-decreasing and communications are FIFO,
no monitor can send a new observation with global time below \(\tmin_1-\sk\).
Thus, we know that all events of global time below \(\tmin_1-\sk\) have already been collected by $M_i$.
The set of possible traces of the system corresponding to that observation is 
$\atw(\tw_i(t))_{\mid \tmin_1-\sk}$.
This is the purpose of the first part of the following proposition.
The second part claims that, indeed, the observation at $\tmin_1(t)-\sk$
is in the set of ``tubes'' of possible observations of $\tw^g$ restricted to $\tmin_1(t)-\sk$.
\begin{restatable}{proposition}{tminone}
  \label{prop-tmin1}
  For any monitor $M_i$, at any local time $t\geq\sk$, 
  \(\tw^g_{\mid \tmin_1(t)-\sk}\) belongs to \(\nolang{\atw(\tw_i(t))_{\mid \tmin_1(t)-\sk}}\).
  Similarly, \(\tw_i(t)_{\mid \tmin_1(t)-\sk}\) belongs to \(\nolang{\atw(\tw^g)_{\mid \tmin_1(t)-\sk}}\).
\end{restatable}
\subsection{Verdicts at $\tmin_1 -\sk$}
\label{sub:verdict_tmin}

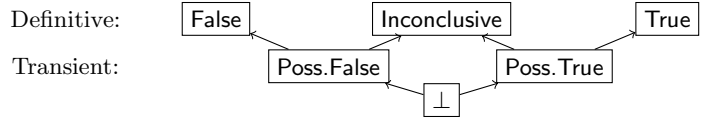
\begin{figure}[t]
\centering
\begin{tikzpicture}[yscale=.9]
  \node[draw] (false) at  (0,0) {\False};
  \node[draw] (inconc) at (3,0) {\Inconc};
  \node[draw] (true) at  (6,0) {\True};
  \node[draw] (pfalse) at (1.5,-.7) {\PFalse};
  \node[draw] (ptrue) at  (4.5,-.7) {\PTrue};
\path[use as bounding box] (3,-1.2);
  \node[draw] (bot) at    (3,-1.2)     {\Pending};
  \draw[->] (bot) edge (pfalse)
            (bot) edge (ptrue)
            (pfalse) edge (false)
            (pfalse) edge (inconc)
            (ptrue) edge (true)
            (ptrue) edge (inconc)
  ;
  \node (definitive) at (-2,0) {Definitive:};
  \node (transient) at (-2,-.7) {Transient:};
\end{tikzpicture}
\caption{Hasse diagram of the verdict preorder} 
\label{fig:verdicts}
\end{figure}
As demonstrated by \cref{prop-tmin1},
at global time $\tmin_1-\sk$ the global trace \(\tw^g_{\mid \tmin_1-\sk}\) necessarily belongs to the language of traces compatible with the collected trace for $M_i$ at local time~$t$, restricted to $\tmin_1-\sk$.
Consequently, this~is a sufficient language to ensure safe monitoring, in~the sense that a \emph{definitive} verdict (\True, \False or \Inconc) for this language built from a current
 observed trace cannot be changed by future observations. 

To go a step further,
we add \emph{transient} verdicts  to the definitive verdicts: 
\PTrue (resp. \PFalse) means that the property can not be falsified (resp. verified), although it is still possible to reach an \Inconc verdict instead of a~\True (resp.~\False)~one. 
The induced pre-order of verdicts \(\lesssim\) is displayed in \cref{fig:verdicts}.
The~{\em verdict function}~$V(t)$ then links 
at each date $t$ the compatible traces $\nolang{\atw(\tw_i(t))_{\mid \tmin_1(t)-\sk}}$ with the property specified by $\ta_i$. Formally
$V(t)$ is
 \begin{itemize}
 \item \(\True\) when $\nolang{\atw(\tw_i(t))_{\mid \tmin_1(t)-\sk}}$ is a good language prefix;
\item \(\False\) when $\nolang{\atw(\tw_i(t))_{\mid \tmin_1(t)-\sk}}$ is a bad language prefix;
\item \(\Inconc\) when
  $\nolang{\atw(\tw_i(t))_{\mid \tmin_1(t)-\sk}}$
  intersects both languages $\sem{\ta_i,\Inev(\Final_i)}$
  and $\sem{\ta_i,\Never(\Final_i)}$;
\item \(\PTrue\) when $\nolang{\atw(\tw_i(t))_{\mid \tmin_1(t)-\sk}}$ intersects $\sem{\ta_i,\Inev(\Final_i)}$ but no higher verdict (\True or \Inconc) applies; 
\item \(\PFalse\) when $\nolang{\atw(\tw_i(t))_{\mid \tmin_1(t)-\sk}}$ intersects $\sem{\ta_i,\Never(\Final_i)}$ but no higher verdict (\False or \Inconc) applies; 
\item \(\Pending\) by default, if  none of these conditions holds.
\end{itemize}
 Because  properties are specified by complete TAs, all traces and their continuations are traces of the TA,
and since good\slash bad prefixes are closed under continuation, 
verdicts can only progress to higher verdicts in the above pre-order:
\begin{restatable}{lemma}{verdictprogress}
  \label{lm:verdict_progress}
  For a fixed monitor \(M_i\) with a property \((\ta_i,\Final_i)\), for any $t'>t$, we have $V(t)\lesssim V(t')$.
\end{restatable}  
In particular, \emph{definitive} verdicts remain true eternally as soon as they hold. 
We~say that \(\tw\)~is \emph{\(\sk\)-conclusive} when \(\atw(\tw)\) yields a definitive non-\Inconc verdict (recall that \(\atw(\tw)\) depends on \sk for the size of its intervals). 
Equivalently: 
\begin{definition}
For a fixed monitor \(M_i\) with a property \((\ta_i,\Final_i)\) and a skew \(\sk\), we say that a timed word \(\tw\) is \emph{\(\sk\)-conclusive} when \(\nolang{\atw(\tw)}\) is either a good or a bad language prefix. 
\end{definition}

\subsection{Data structure}
\label{sub:CS}
So far, we have defined verdicts with respect to the language
\(\nolang{\atw(\tw_i(t))_{\mid\tmin_1(t)-\sk}}\).  However, in order
to incrementally compute verdicts when new observations arrive,
it~will be more adequate to manipulate sets of configurations reached
after this language.
The next proposition justifies the adequacy of this approach.

\begin{restatable}{proposition}{statecheck}
  \label{pr:statecheck}
  The verdict $V(t)$ at $M_i$-local date $t$ can be computed by checking
  whether
    the set of configurations \({(\initloc,\initv,0)}\, \after_0^{\tmin_1(t)-\sk}\, \nolang{\atw(\tw_i(t))_{\mid\tmin_1(t)-\sk}}\) intersects or is included in 
\(\Inev(\Final_i)\) or \(\Never(\Final_i)\).
\end{restatable}

\looseness=-1
In the following, we proceed in steps to define the computation of the above set of configurations: first, we decompose the computation on \emph{unordered} languages into computations on \emph{ordered~ones}, where permutations of events are fixed (\cref{cr:decomp_ordlang}).
We~encode them in a
data structure~$\CS(t)$
that represents both the set of configurations reached after some trace in \(\nolang{\atw(\tw_i(t))_{\mid\tmin_1(t)-\sk}}\),
  and the remaining events necessary for incremental computation. 
Then, we~intersect the sets of configurations with the precise time of interest ($\tmin_1-\sk$) to obtain the equality with the above set (\cref{cr:R_decomp}).
Finally, we show how to incrementally update the proposed data structure when new observations are collected.

\paragraph{Decomposition of \(\{(\initloc,\initv,0)\}\, \after\, \nolang{\atw(\tw_i(t))_{\mid \tmin_1(t)-\sk}}\).} 
The~separation of an unordered language 
into a union  of ordered ones
requires considering permutations of the restriction \(\atw(\tw_i(t))_{\mid \tmin_1(t)-\sk}\).
For~this, we define the \emph{decomposition} of an ATW~$\atw$ at some date~$T$,
as the set of all possible prefixes of~$\atw_{\mid T}$
paired with
the set of events that are yet to be accounted for, as this set will be necessary to incrementally update the data structure.

\begin{definition}
  The \emph{decomposition of $\atw \in \ATW(\Act)$ at global time $T$} is 
\[\Decomp(\atw,T)=\{(f({\atw_1}) ; {\atw_2})\mid
\atw= \atw_1\tens \atw_2 \wedge
\atw_1 \preceq_{[0,T]} \atw \wedge 
f\in\calF(\atw_1)\}.
\]
\end{definition} 

For a pair \(({\atw_1};{\atw_2})\) in $\Decomp(\atw,T)$,  \({\atw_1}\)
corresponds to a permutation of an element in~\(\atw_{\mid T}\),
while \({\atw_2}\) lists the remaining events that are not taken into account in~\(\atw_1\).
We can then use the decomposition to express the unordered language of a restriction as a union of ordered ones as follows. 
\begin{restatable}{proposition}{decompordlang}
  \label{cr:decomp_ordlang}
    \(\nolang{\atw_{\mid T}}=\bigcup_{(\atw_1;\atw_2)\in\Decomp(\atw,T)}\ordlang{{\atw_1}_{\cap[0,T]}}\)
\end{restatable}

We are now ready to define our data structure \(\CS(t)\)  (we~should write \(\CS_i(t)\) but the monitor is~clear from the context), which encodes the configurations reached through each~\(\atw_1\) in the decomposition, associated with its remainder~\(\atw_2\):
\begin{definition}
  For a monitor~$M_i$ and a local time~$t$, we~define \(\CS(t)\), as 
  \[
  \CS(t)= \bigl\{(\{(\initloc,\initv,0)\}\, \after \, \ordlang{\atw_1};
  \atw_2)
  \bigm|
  (\atw_1;\atw_2)\in\Decomp(\atw(\tw_i(t)),\tmin_1(t)-\sk)
  \bigr\}.
  \]
\end{definition}

The~set~\(\CS(t)\) represents a set of configurations in which the system can be
after some sequence of events in~\(\atw(\tw_i(t))_{\mid \tmin_1(t)-\sk}\).
A~restriction to dates before $\tmin_1(t)-\sk$ is still necessary 
(see~the intersection with~\([0,T]\) 
in \cref{cr:decomp_ordlang}).
In order to~add this constraint, we~call \emph{state of $\CS(t)$} the set of configurations
\(\state(\CS(t))=\bigcup_{(S,\atw_2)\in\CS(t)}
\{(\loc,\val,t)\in S\mid t=\tmin_1(t)-\sk\}\) 
and get the desired equality:
\begin{restatable}{proposition}{Rdecomp}
  \label{cr:R_decomp}
  \(\state(\CS(t))=  \{(\initloc,\initv,0)\}\, \after_0^{\tmin_1(t)-\sk}\, \nolang{\atw(\tw_i(t))_{\mid \tmin_1(t)-\sk}}\)
\end{restatable}

\begin{example}
  Let us consider the example from \cref{fig-extmin} with the collected trace at time $t=10$ being $\sigma_i(t)=({\color{red} a}, 1)({\color{green!40!black} c}, 2)({\color{blue} b}, 3)({\color{blue} b}, 5)({\color{green!40!black} c}, 5.5)({\color{red} a}, 7)({\color{red} a}, 10)$ and the skew $\sk=0.7$. We have $\tmin_1-\sk= 4.8$. 
We first have to consider the set $C$  of all possible configurations reached by an interleaving of the first three observations $({\color{red} a}, 1)({\color{green!40!black} c}, 2)({\color{blue} b}, 3)$. 
Then, we have to consider every case for the events $({\color{blue} b}, 5)({\color{green!40!black} c}, 5.5)$ that can occur before or after $\tmin_1-\sk$. $\CS(10)$ is then composed of the following elements, for each $c\in C$: 
\begin{itemize}
\item $(c ; \atw(({\color{blue} b}, 5)({\color{green!40!black} c}, 5.5)({\color{red} a}, 7)({\color{red} a}, 10)))$, meaning that %
  all events occur after time~$4.8$;
\item $(c\, \after \, \ordlang{({\color{blue} b}, [4.3, 5.7])} ; \atw(({\color{green!40!black} c}, 5.5)({\color{red} a}, 7)({\color{red} a}, 10)))$, considering that the event $b$ happened before time $4.8$;
\item $(c\, \after \, \ordlang{({\color{green!40!black} c}, [4.8, 6.2])} ; \atw(({\color{blue} b}, 5)({\color{red} a}, 7)({\color{red} a}, 10)))$, considering that $c$ occured before time $4.8$.
\item the two elements $(c\, \after \, \ordlang{({\color{blue} b}, [4.3, 5.7])({\color{green!40!black} c}, [4.8, 6.2])} ;\atw(({\color{red} a}, 7)({\color{red} a}, 10)))$ and 
$(c\, \after \, \ordlang{({\color{green!40!black} c}, [4.8, 6.2])({\color{blue} b}, [4.3, 5.7])} ;\atw(({\color{red} a}, 7)({\color{red} a}, 10)))$  considering that both $b$ and $c$ occurred before time~$4.8$, thus both orderings should be considered.
\end{itemize}

\looseness=-1
Note that
$\CS(10)$
contains configurations that can only be reached if the collected events occur after $\tmin_1-\sk$, but they do not appear in $\state(\CS(10))$, since we only consider configurations that are reached at time $\tmin_1-\sk$, meaning that the events leading to these configurations must have occurred before this time.
\end{example}

\paragraph{Updates of \(\CS\).} \looseness=-1
When time passes, new events may be collected.
If~they do not change~\(\tmin_1\), the~only updates to \(\CS\) is their addition to~\(\atw_2\). 
If $\tmin_1$ changes at~$t'>t$ because of newly collected events~$\atw'$, then the combination of \(\CS(t)\) and~$\atw'$ 
contains all the necessary information to update~$\CS$, as each element of~$\CS(t)$ encodes all of \(\atw(\tw_i(t))_{\mid \tmin_1(t)-\sk}\).
Thus, updating the structure is only a matter of selecting, for each \((S,\atw_2) \in \CS(t)\), the~possible sub-words of \(\atw_2\cdot\atw'\) to apply after~$S$. %

\begin{restatable}{proposition}{updateCS}
  \label{pr:update_CS}  
  Let $t' \geq t$ and $\atw'$ be the sequence of events received in
  the interval $(t,t']$ (\ie, such that $\tw_i(t') = \tw_i(t) \tens
    \atw'$), then
    \[
    \CS(t')=
    \bigl\{ (S \; \after \; \ordlang{\atw'_1}; \atw'_2)
    \bigm| %
     (S; \atw_2)\in \CS(t),\ 
     (\atw_1',\atw_2')\in\Decomp(\atw_2\tens\atw',\tmin_1(t')-\sk)
    \bigr\}.
    \]
\end{restatable}
Intuitively, for each pair $(S,\atw_2)$ in $\CS(t)$, 
the extension of $\atw_2$ with the newly collected events $\atw'$ is decomposed at $\tmin_1(t')-\sk$.
For each possible element  $(\atw'_1,\atw'_2)$ in this decomposition,
\(\CS(t')\) builds the pair  made of the set $S \; \after \; \ordlang{\atw'_1}$ associated with the remainder $\atw'_2$.
We call \(next(\CS(t),t')\) the function that computes $\CS(t')$ at the time \(t'\) when \(\tmin_1(.)\) changes according to \cref{pr:update_CS}. 
We~now have a data structure that can be used to compute the set of configurations needed to infer verdicts (\cref{cr:R_decomp}) and can be updated incrementally based on the new collected observations and \(\tmin_1(\cdot)-\sk\) (\cref{pr:update_CS}).

\subsection{Monitoring at $\tmin_1(.) -\sk$}
\label{sub:monit_tmin}
\looseness=-1
The previous discussions  lead to the monitoring algorithm presented in \cref{alg:tmin_monit}.
It uses a triple of boolean values $(I,N,C)$ encoding the intersection of \(\state(\CS)\) respectively with
$\Inev(\Final_i)$, $\Never(\Final_i)$ and the complement of their union.
The algorithm starts with \(\CS\) being the initial state with no remaining events, the minimal time for monitoring \(\tmin=\sk\),
\(\jtmin\) initially set to $(k,0)_{k\in \IntInv{n}}$.
Each new collected sequence of observations from a monitor $M_j$  (recall that monitors group all events with same date in a unique message) is added to all continuations \(\atw\) in $\CS$,  and \(\jtmin\) is updated.
If~\(\tmin_1\) has changed, %
$\tmin$~and \(\CS\) %
are updated (as~discussed above).
The~update of~$(I,N,C)$ determines the verdict which is returned if definitive
(a~lazy evaluation of~$(I,N,C)$ optimizes the update, $I$~and $N$ being non-decreasing, while $C$ is non-increasing). We do not detail here the communication of verdicts between monitors which could help anticipate their termination.

\begin{algorithm}
  \SetAlgoLined
  \SetKw{kwRecv}{Receive}
\kwInit{\( \CS=\{((\initloc,\initv,0);[])\}\); \(\tmin=\sk\); \(\jtmin=(k,0)_{k \in\IntInv{n}}\);
 \(\verdict(i):=\Pending\); 
 \((I,N,C):= ((\initloc,\initv,0)\in \Inev(\Final_i),(\initloc,\initv,0)\in \Never(\Final_i), \neg(I\vee N))\);
}
\While {True} {
  \kwRecv{sequence \((a_1,t_a)\dots(a_n,t_a)\)  from monitor \(M_j, j \in\IntInv{n}\)}\;
\(\CS:=\{(S,\atw\tens\atw((a_1,t_a)\dots(a_n,t_a)))\mid (S,\atw)\in\CS\}\)\;
 update $\jtmin$ \;
\If{\(\tmin_1>\tmin\)}{ %
    \(\tmin:=\tmin_1\)\;
    \(\CS := next(\CS, \tmin_1-\sk) \) \; %
         $I:= I \,\vee \, (\state(\CS) \cap  \Inev(\Final_i)  \neq \emptyset)$;\\
        $N:= N \,\vee\,  (\state(\CS) \cap  \Never(\Final_i)  \neq \emptyset)$;\\
       $C:= C\,\wedge\, (\state(\CS) \cap  \overline{\Inev(\Final_i) \cup \Never(\Final_i)})  \neq \emptyset$; \\
    \Switch{$(I,N,C)$}{ 
    \lCase {$(1,1,*)$} {\return(\verdict(i):=\Inconc)} 
    \lCase {$(1,0,0)$} {\return(\verdict(i):=\True) }
    \lCase {$(0,1,0)$} {\return(\verdict(i):=\False)}
    \lCase {$(1,0,1)$} {\verdict(i):=\Poss.\True }
    \lCase {$(0,1,1)$} {\verdict(i):=\Poss.\False}
    }
}
} 
\caption{The monitor $M_i$'s algorithm to monitor at $\tmin_1(.)-\sk$.}
\label{alg:tmin_monit}
\end{algorithm}

\begin{restatable}{proposition}{maxverdict}
  \label{pr:maxverdict}
\Cref{alg:tmin_monit} sets the verdict to $V(t)$. 
\end{restatable}

We can prove the following soundness and completeness of the monitoring algorithm. Notice that completeness is limited to \(2\sk\)-conclusive executions.
\begin{restatable}{theorem}{soundcomplete}\label{thm:tmin_monit}
  Monitoring at $\tmin_1(.)-\sk$ is sound and complete where, for any local monitor \(M_i\) and
   its property~\((\ta_i,\Final_i)\):
\begin{description}
\item[soundness] means that for any global trace \(\tw^g\in \TT(\Act)\) produced at date $T\in \Rplus$,
  if \(\verdict(i) = \True\) at time~$T$, then $\tw^g$ is a good prefix of \((\ta_i,\Final_i)\)
  (respectively, if~\(\verdict(i) = \False\) at time~$T$, then $\tw^g$ is a bad prefix of \((\ta_i,\Final_i)\)).
   Furthermore, if \(\verdict(i)=\Inconc\), then neither $\tw^g$ nor its possible continuations are \(2\sk\)-conclusive on \((\ta_i,\Final_i)\). 
\item[completeness] means that for any global trace \(\tw^g\in\TT(\Act)\), 
  if a prefix %
  of $\tw^g$ is \(2\sk\)-conclusive and good (respectively, bad) on \((\ta_i,\Final_i)\) %
  then there exists 
  some date~${T \in \Rplus}$
  such that  \(\verdict(i) = \True\) 
  (respectively,  \(\verdict(i)= \False\)).
\end{description}
\end{restatable}

%% file: conclu.tex
\section{Conclusion}
This paper presents a distributed approach to monitor properties specified as deterministic timed automata when faced with approximation on events dates. The approach relies on the identification of the 
point in time at which sufficient information has been gathered by the local monitor to compute a verdict and 
the incremental computation of the set of states
of the property compatible with the collected observation at this point in time
This requires the careful account of potential permutations of events emanating from distant components.

This method allows to apply monitoring on complex systems that are \emph{distributed} in space and whose behaviours \emph{depend strongly on time}, further increasing the reach of this popular runtime verification method. It is interesting to notice the timely nature of this contribution, as the interest for distributed systems is developing not only in verification (see related works) but also in model learning (\cite{NS23,LGHM23}~discrete time), which could soon allow to automatically generate models of systems and specifications, allowing---at longer term---for fully black-box distributed tools requiring no expert knowledge. 

While we ensured the soundness and completeness of our algorithm, its efficiency still needs to be experimented.
Future works should include the implementation and test of this algorithm against realistic models and properties. This implementation could be then compared to the one presented in~\cite{DBLP:conf/time/RoussanalyF22} that allows decentralized monitoring of regular timed expressions (but without clock skew), which can be generated from a timed automaton.
Additionally, our algorithm could be extended in several ways.
First, with additional hypotheses (\eg, maximal throughput of components), we could issue verdicts earlier by anticipating the occurrence of events.
This would require an extension of the structure $\CS$, adding a level of uncertainty.
The balance between the gain of anticipation and the cost of updating this structure would certainly be an issue and requires experimental tuning. 
Using similar techniques,
we~believe we can handle properties defined by non-deterministic timed automata, at~least for one-clock automata~\cite{abs-2002-07049}.
Finally, we~can also try to reduce the communication overhead.
Indeed, we assumed that all the local observations are forwarded to every other monitor, which can be improved in several ways, depending on the system topology.

%% file: main.bbl
\begin{thebibliography}{10}

\bibitem{ALUR1994183}
Rajeev Alur and David~L. Dill.
\newblock A theory of timed automata.
\newblock {\em Theoretical Computer Science}, 126(2):183--235, 1994.

\bibitem{AsarinCM02}
Eugene Asarin, Paul Caspi, and Oded Maler.
\newblock Timed regular expressions.
\newblock {\em J. {ACM}}, 49(2):172--206, 2002.

\bibitem{BK08}
{\relax Ch}ristel Baier and Joost-Pieter Katoen.
\newblock {\em Principles of Model-Checking}.
\newblock MIT Press, May 2008.

\bibitem{BF18}
Ezio Bartocci and Yli{\`e}s Falcone, editors.
\newblock {\em Lectures on Runtime Verification}, volume 10457 of {\em Lecture
  Notes in Computer Science}.
\newblock Springer-Verlag, 2018.

\bibitem{Basin_2019}
David Basin, Felix Klaedtke, and Eugen Zălinescu.
\newblock Runtime verification over out-of-order streams.
\newblock {\em ACM Transactions on Computational Logic}, 21(1):1–43, October
  2019.

\bibitem{BauerLS11}
Andreas Bauer, Martin Leucker, and Christian Schallhart.
\newblock Runtime verification for {LTL} and {TLTL}.
\newblock {\em ACM Trans. Softw. Eng. Methodol.}, 20(4):14:1--14:64, September
  2011.

\bibitem{BauerF12}
Andreas~Klaus Bauer and Yli{\`{e}}s Falcone.
\newblock Decentralised {LTL} monitoring.
\newblock In Dimitra Giannakopoulou and Dominique M{\'{e}}ry, editors, {\em
  {FM} 2012: Formal Methods - 18th International Symposium, Paris, France,
  August 27-31, 2012. Proceedings}, volume 7436 of {\em Lecture Notes in
  Computer Science}, pages 85--100. Springer, 2012.

\bibitem{BY04}
Johan Bengtsson and Wang Yi.
\newblock Timed automata: Semantics, algorithms and tools.
\newblock In J{\"o}rg Desel, Wolfgang Reisig, and Grzegorz Rozenberg, editors,
  {\em Lectures on Concurrency and {P}etri Nets}, volume 2098 of {\em Lecture
  Notes in Computer Science}, pages 87--124. Springer-Verlag, 2004.

\bibitem{DBLP:journals/sttt/BouyerHJJM21}
Patricia Bouyer, L{\'{e}}o Henry, Samy Jaziri, Thierry J{\'{e}}ron, and Nicolas
  Markey.
\newblock Diagnosing timed automata using timed markings.
\newblock {\em Int. J. Softw. Tools Technol. Transf.}, 23(2):229--253, 2021.

\bibitem{hbmc}
Edmund~M. Clarke, Thomas~A. Henzinger, Helmut Veith, and Roderick Bloem.
\newblock {\em Handbook of Model Checking}.
\newblock Springer-Verlag, April 2018.

\bibitem{ColomboF16}
Christian Colombo and Yli{\`{e}}s Falcone.
\newblock Organising {LTL} monitors over distributed systems with a global
  clock.
\newblock {\em Formal Methods Syst. Des.}, 49(1-2):109--158, 2016.

\bibitem{El-HokayemF20}
Antoine El{-}Hokayem and Yli{\`{e}}s Falcone.
\newblock On the monitoring of decentralized specifications: Semantics,
  properties, analysis, and simulation.
\newblock {\em {ACM} Trans. Softw. Eng. Methodol.}, 29(1):1:1--1:57, 2020.

\bibitem{Falcone21}
Yli{\`{e}}s Falcone.
\newblock On decentralized monitoring.
\newblock In Ayoub Nouri, Weimin Wu, Kamel Barkaoui, and ZhiWu Li, editors,
  {\em Verification and Evaluation of Computer and Communication Systems - 15th
  International Conference, VECoS 2021, Virtual Event, November 22-23, 2021,
  Revised Selected Papers}, volume 13187 of {\em Lecture Notes in Computer
  Science}, pages 1--16. Springer, 2021.

\bibitem{FalconeCF14}
Yli{\`{e}}s Falcone, Tom Cornebize, and Jean{-}Claude Fernandez.
\newblock Efficient and generalized decentralized monitoring of regular
  languages.
\newblock In Erika {\'{A}}brah{\'{a}}m and Catuscia Palamidessi, editors, {\em
  Formal Techniques for Distributed Objects, Components, and Systems - 34th
  {IFIP} {WG} 6.1 International Conference, {FORTE} 2014, Held as Part of the
  9th International Federated Conference on Distributed Computing Techniques,
  DisCoTec 2014, Berlin, Germany, June 3-5, 2014. Proceedings}, volume 8461 of
  {\em Lecture Notes in Computer Science}, pages 66--83. Springer, 2014.

\bibitem{Fil11}
Jean-Christophe Filli{\^a}tre.
\newblock Deductive software verification.
\newblock {\em International Journal on Software Tools for Technology
  Transfer}, 13(5):397--403, October 2011.

\bibitem{FrancalanzaPS18}
Adrian Francalanza, Jorge~A. P{\'{e}}rez, and C{\'{e}}sar S{\'{a}}nchez.
\newblock Runtime verification for decentralised and distributed systems.
\newblock In Ezio Bartocci and Yli{\`{e}}s Falcone, editors, {\em Lectures on
  Runtime Verification - Introductory and Advanced Topics}, volume 10457 of
  {\em Lecture Notes in Computer Science}, pages 176--210. Springer, 2018.

\bibitem{abs-2107-06084}
Florian Gallay and Yli{\`{e}}s Falcone.
\newblock Decentralized {LTL} enforcement.
\newblock In Pierre Ganty and Davide Bresolin, editors, {\em Proceedings 12th
  International Symposium on Games, Automata, Logics, and Formal Verification,
  GandALF 2021, Padua, Italy, 20-22 September 2021}, volume 346 of {\em
  {EPTCS}}, pages 135--151, 2021.

\bibitem{GANGULY-JPDC-2024}
Ritam Ganguly, Yingjie Xue, Aaron Jonckheere, Parker Ljung, Benjamin
  Schornstein, Borzoo Bonakdarpour, and Maurice Herlihy.
\newblock Distributed runtime verification of metric temporal properties.
\newblock {\em Journal of Parallel and Distributed Computing}, 185:104801,
  2024.

\bibitem{abs-2002-07049}
Alejandro Grez, Filip Mazowiecki, Michal Pilipczuk, Gabriele Puppis, and
  Cristian Riveros.
\newblock The monitoring problem for timed automata.
\newblock {\em CoRR}, abs/2002.07049, 2020.

\bibitem{Hoa69}
Charles Antony~Richard Hoare.
\newblock An axiomatic basis for computer programming.
\newblock {\em Communications of the ACM}, 12(10):576--580, October 1969.

\bibitem{KohlHermanns23}
Maximilian~A. K{\"{o}}hl and Holger Hermanns.
\newblock Model-based diagnosis of real-time systems: Robustness against
  varying latency, clock drift, and out-of-order observations.
\newblock {\em {ACM} Transactions on Embedded Computing Systems},
  22(4):68:1--68:48, 2023.

\bibitem{LGHM23}
Faezeh Labbaf, Jan~Friso Groote, Hossein Hojjat, and Mohammad~Reza Mousavi.
\newblock Compositional learning for interleaving parallel automata.
\newblock In Orna Kupferman and Pawel Sobocinski, editors, {\em Foundations of
  Software Science and Computation Structures - 26th International Conference,
  FoSSaCS 2023, Held as Part of the European Joint Conferences on Theory and
  Practice of Software, {ETAPS} 2023, Paris, France, April 22-27, 2023,
  Proceedings}, volume 13992 of {\em Lecture Notes in Computer Science}, pages
  413--435. Springer, 2023.

\bibitem{LS09}
Martin Leucker and Christian Schallart.
\newblock A brief account of runtime verification.
\newblock {\em Journal of Logic and Algebraic Programming}, 78(5):293--303, May
  2009.

\bibitem{NS23}
Thomas Neele and Matteo Sammartino.
\newblock Compositional automata learning of synchronous systems.
\newblock In {\em International Conference on Fundamental Approaches to
  Software Engineering}, pages 47--66. Springer Nature Switzerland, 2023.

\bibitem{NickovicM07}
Dejan Nickovic and Oded Maler.
\newblock {AMT}: a property-based monitoring tool for analog systems.
\newblock In Jean-Fran\c{c}ois Raskin and P.~S. Thiagarajan, editors, {\em
  Proceedings of the 5th International Conference on Formal modeling and
  analysis of timed systems (FORMATS 2007)}, volume 4763 of {\em Lecture Notes
  in Computer Science}, pages 304--319. Springer-Verlag, 2007.

\bibitem{PinisettyJTFMP17}
Srinivas Pinisetty, Thierry J{\'{e}}ron, Stavros Tripakis, Yli{\`{e}}s Falcone,
  Herv{\'{e}} Marchand, and Viorel Preoteasa.
\newblock Predictive runtime verification of timed properties.
\newblock {\em J. Syst. Softw.}, 132:353--365, 2017.

\bibitem{DBLP:conf/time/RoussanalyF22}
Victor Roussanaly and Yli{\`{e}}s Falcone.
\newblock Decentralised runtime verification of timed regular expressions.
\newblock In Alexander Artikis, Roberto Posenato, and Stefano Tonetta, editors,
  {\em 29th International Symposium on Temporal Representation and Reasoning,
  {TIME} 2022, November 7-9, 2022, Virtual Conference}, volume 247 of {\em
  LIPIcs}, pages 6:1--6:18. Schloss Dagstuhl - Leibniz-Zentrum f{\"{u}}r
  Informatik, 2022.

\bibitem{ThatiR05}
Prasanna Thati and Grigore Rosu.
\newblock Monitoring algorithms for metric temporal logic specifications.
\newblock {\em Electronic Notes in Theoretical Computer Science}, 113:145--162,
  2005.

\bibitem{Ulus17}
Dogan Ulus.
\newblock Montre: {A} tool for monitoring timed regular expressions.
\newblock In Rupak Majumdar and Viktor Kuncak, editors, {\em Computer Aided
  Verification - 29th International Conference, {CAV} 2017, Heidelberg,
  Germany, July 24-28, 2017, Proceedings, Part {I}}, volume 10426 of {\em
  Lecture Notes in Computer Science}, pages 329--335. Springer, 2017.

\bibitem{WangYL07}
Yin Wang, Tae{-}Sic Yoo, and St{\'{e}}phane Lafortune.
\newblock Diagnosis of discrete event systems using decentralized
  architectures.
\newblock {\em Discret. Event Dyn. Syst.}, 17(2):233--263, 2007.

\bibitem{YinL15}
Xiang Yin and St{\'{e}}phane Lafortune.
\newblock Codiagnosability and coobservability under dynamic observations:
  Transformation and verification.
\newblock {\em Autom.}, 61:241--252, 2015.

\end{thebibliography}
